\documentclass[12pt,preprint]{aastex}

\shorttitle{\text{Swift} Long Lag GRBs}
\shortauthors{Xiao and Schaefer}

\begin{document}


\title{Are \textit{Swift} Long-Lag Gamma-Ray Bursts in the Local Supercluster?}


\author{Limin Xiao and Bradley E. Schaefer}
\affil{Physics and Astronomy, Louisiana State University, Baton Rouge, LA, 70803}

\email{lxiao1@lsu.edu}






\begin{abstract}
A sample of 18 long-lag ($\tau_{lag} > 1$ s) Gamma-Ray Bursts (GRBs) has been drawn from our catalog of all \textit{Swift} long GRBs. Four different tests are done on this sample to test the prediction that a large fraction of long-lag GRBs are from our Local Supercluster. The results of these four tests come out that: (1) the distribution of these GRBs shows no tendency towards the Supergalactic plane; (2) the distribution shows no tendency towards the Virgo or Coma Cluster; (3) no associated bright host galaxies ($m\leqslant15$) in the Local Supercluster are found for any of the 18 GRBs; (4) 17 of these 18 GRBs have redshifts of $z>0.5$, which are too far to be in the Local Supercluster. All these results disproved the hypothesis that any significant fraction of long-lag GRBs are from Local Supercluster. Hence these long-lag GRBs can not be counted in the calculation of LIGO detection rates. An explanation of why we can detect long-lag GRBs at high redshift is presented.
\end{abstract}



\section{Introduction}

Gamma-Ray Bursts (GRBs) are bursts of gamma radiations that are isotropically distributed on the sky. Their time duration $T_{90}$ ranges from $\sim0.1$ s up to $\sim1000$ s, and their measured spectroscopic redshifts range is $0.008<z< 6.7$. Based on the time duration, they are divided into two different groups: GRBs with $T_{90}<2$ s are classified as short duration bursts, and those with $T_{90}>2$ s as long duration GRBs. The spectral lag ($\tau_{lag}$) of a GRB is a parameter that measures the delay time between the soft and hard light curves of the GRB. The global hard-to-soft spectral evolution of GRB pulses was found by \citet{nor86} in analysis of SMM (Solar Maximum Mission satellite) GRB data, and a cross-correlation analysis between different channels of data for calculating $\tau_{lag}$ values was performed by \citet{ban97}. A power law function between the $\tau_{lag}$ value and the peak luminosity (L) was well fitted for six BATSE and BeppoSAX long GRBs with measured redshifts \citep{nor00}. In the power law function, the $\tau_{lag}$ is corrected for the cosmological time dilation effect by dividing a factor of (1+z). It was also pointed out in the same paper that GRB980425 with a long $\tau_{lag}$ value ($\tau_{lag}=2.8$ s) falls far below the power law fitting curve by a factor of several hundred. The empirical $\tau_{lag}-L$ relation can be simply explained as a consequence of radiative cooling of the shocked material in the jet  \citep{sch04}. High-luminosity bursts will have fast radiative cooling and hence short lags, while low-luminosity bursts will have slow radiative cooling and hence long lags. This general result predicts that the burst luminosity should be proportional to $\tau_{lag}^{-1}$ and that is exactly what is observed. 

The $\tau_{lag}$ analysis on BATSE and INTEGRAL samples shows a distribution from $\sim0 - 10$ s for long GRBs \citep{nor02, fol08}, with most of these $\tau_{lag}$ concentrated in the 0 - 1 s region. According to the $\tau_{lag}-L$ relation, a long $\tau_{lag}$ corresponds with a low luminosity, and for a low luminosity GRB to be detected by our instruments, it should be relatively nearby. \citet{nor02} pointed out that GRB980425 might represent a subclass of long GRBs, with long $\tau_{lag}$, soft spectrum, ultra-low luminosity, and nearby. In this case, a possible break might exist in the $\tau_{lag}-L$ relation in the long $\tau_{lag}$ region, which would indicate that these long $\tau_{lag}$ GRBs are even closer than what is predicted by the $L\propto \tau_{lag}^{-1}$ relation. Indeed, two long $\tau_{lag}$ bursts are confidently known to be at distances close enough to be inside the Local Supercluster. GRB980425, with $\tau_{lag}=2.8$ s, had an ultra-low luminosity, and lies in a galaxy only $\sim38$ Mpc away \citep{gal98}. GRB830801, is the all-time brightest GRB yet has a long $\tau_{lag}$ ($2.2\pm0.2$ s), so a very low redshift of $z\sim0.01$ is calculated from the $\tau_{lag}-L$ relation \citep{sch01}. GRB830801 also happens to be from a direction close to the Virgo Cluster.

Given that these long $\tau_{lag}$ GRBs might be nearby, is there any local structure of galaxies to host these GRBs? The Local Supercluster was proposed by \citet{dev53}, from an investigation of spatial distribution of galaxies. It was first named as `Supergalaxy', which was later changed to be `Local Supercluster' \citep{dev58}. More detailed studies show that the main body of the Local Supercluster is a filamentary structure extending over $\sim40~h^{-1}$ Mpc, and is centered on the Virgo Cluster \citep{tul87, kar96, lah00}. Around $60\%$ of the luminous galaxies in  the volume of Local Supercluster are within the structure that defines the plane of the Supercluster ($20\%$ in Virgo Cluster and $40\%$ in Virgo II Cloud and Canes Venatici Cloud), and most of the remaining $40\%$ lies within five clouds off the plane, which is called a `halo' \citep{tul82}. Our Local Group is in the outskirt of this region. 

\citet{nor02} presented a catalog of $\tau_{lag}$ values for 1429 BATSE long GRBs, from which a sample of 64 long $\tau_{lag}$ GRBs (with $\tau_{lag}>2$ s) was selected. These $\tau_{lag}$ values were calculated in the observer's rest frame, without making the time dilation correction (which should be small for local bursts). By plotting these long $\tau_{lag}$ bursts on a sky map in Supergalactic coordinates, a concentration towards the Supergalactic plane was found, with three-fourth of these bursts located in the half of the sky between $-30^{\circ}$ and $30^{\circ}$ of Supergalactic latitude. Quantitatively, the quadruple moment of these GRBs is roughly $-0.10 \pm 0.04$, which shows a $2.5\sigma_Q$ deviation from isotropy. This result implies that long $\tau_{lag}$ value will be an indicator for local GRBs. From the solid long GRB-SN connection (e.g. GRB980425 \& SN1998bw, GRB030329 \& SN2003dh) and the model of massive SN (from the collapsing in highly non-axisymmetric modes), strong gravitational waves can be produced at a rate of $\sim4~yr^{-1}$, and these gravitational waves might be able to be detected by LIGO \citep{nor03}.

An independent catalog of long $\tau_{lag}$ bursts discovered by INTEGRAL has been created by \citet{fol08}, with 11 long $\tau_{lag}$ ($\tau_{lag} > 0.75$ s) GRBs being pulled out from the whole INTEGRAL sample. They found that 10 of the 11 long $\tau_{lag}$ bursts are located within the $-30^{\circ}$ to $30^{\circ}$ Supergalactic latitude region. The quadruple moment of the 11 GRBs is $-0.225\pm0.009$. This result is confirmed by \citet{via08}. By comparing with the simulation based on INTEGRAL sky coverage, the quadruple moment is $Q=-0.271\pm0.089$ for long $\tau_{lag}$ GRBs and $Q=-0.007\pm0.042$ for the whole sample. The INTEGRAL result is broadly consistent with the conclusion of \citet{nor02}, however, in \citet{nor02}, the quadruple moments for the samples of $\tau_{lag}>0.5$ s and $\tau_{lag}>1$ s have a substantially lower significance ($Q = -0.022\pm0.020$ for the $\tau_{lag}>0.5$ s sample and $Q = -0.043 \pm 0.026$ for the $\tau_{lag}>1$ s sample). With three results on two independent samples (pointing to a concentration towards the Supergalactic plane), another sample is needed to test the Local Supercluster hypothesis.

\textit{Swift}, the multiwavelength GRB detection satellite, was launched Nov. 2004 \citep{geh04}. It has three instruments on board. The wide field Burst Alert Telescope (BAT), which covers 15 keV to 150 keV energy band, can position a burst to $1'-4'$ accuracy. The narrow X-ray telescope (XRT) and UV/Optical telescope (UVOT) will start observing the GRB within $\sim100$ s after it is triggered and position it within $5''$ and $0.3''$ respectively. Within $\sim100$ s, this accurate position of the GRB will be measured and distributed to the community through GRB Coordinate Network (GCN), and large ground telescopes will be able to follow up and make their own observations. During its four years of operation, \textit{Swift} has been triggered by more than 350 GRBs, $\sim300$ of which have been confirmed as long duration GRBs, and $\sim30\%$ have their spectroscopic or photometric redshift measured. The accurate localizations of \textit{Swift} GRBs make possible the  search for hosts of long $\tau_{lag}$ bursts in Local Supercluster galaxies. The GRB redshift will also directly tell us the distances of these GRBs.

In this paper, we will use \textit{Swift} data to test the hypothesis that most long $\tau_{lag}$ GRBs are in the Local Supercluster. The tests are made of four parts: (1) Is there any tendency of concentration towards the Supergalactic plane? (2) Is there any tendency of concentration towards the Virgo Cluster? (3) Can we find bright host galaxies for these long $\tau_{lag}$ GRBs? (4) Do any of these long $\tau_{lag}$ GRBs have a redshift of $z<0.013$? 

\section{\textit{Swift} Data}

All \textit{Swift} GRB data are available on the Legacy ftp site\footnote{ftp://legacy.gsfc.nasa.gov/swift/}, along with the software published by the \textit{Swift} team. BAT light curves for GRBs within any possible energy bands (15 keV to 150 keV) at any possible time bins ($>0.064$s) can be generated. Conventionally, for the calculation of $\tau_{lag}$, we use the light curves with 0.064 s time bins and energy bands from 25-50 keV and 100-150 keV. By applying a Cross-Correlation Function (CCF) method on the light curves, and fitting the CCF plot with an automatically selected polynomial function, we calculated the $\tau_{lag}$ values for all \textit{Swift} long GRBs. Details of our conventional calculation are presented in \citet{xia08}. In the same paper, by using the $\tau_{lag}$ values as well as four other luminosity indicators (variability, minimum rise time $\tau_{RT}$, number of peaks $N_{peak}$ and peak energy in the spectrum $E_{peak}$), we calculated the redshifts for all \textit{Swift} long GRBs completely independent of spectroscopic redshifts. A comparison between our redshifts, $z_{ind}$, and spectroscopic redshifts $z_{spec}$ ($\chi_{red}^2 = 1.09$, and $<log_{10}[z_{ind}/z_{spec}]> = -0.005 \pm 0.050$) shows that our reported error bars are reasonably good, and our redshift values are not biased, high or low.

From our \textit{Swift} GRB redshift and luminosity indicators catalog (ranging from Dec. 2004 (GRB041220) to Jul. 2008 (GRB080723A)), 18 GRBs with long $\tau_{lag}$ values ($\tau_{lag} > 1$ s) are pulled out. Data for these 18 GRBs are listed in Table 1. Column 1 gives the six digit identification numbers of each GRB. Column 2 listed our measured $\tau_{lag}$ values with their $1-\sigma$ uncertainties. Column 3 gives the spectroscopic redshifts for 6 of these GRBs and our calculated redshifts $z_{ind}$ with $1-\sigma$ uncertainties for the remaining 12. The references for these redshifts are listed in column 4. The celestial right ascension and declination of these GRBs from BAT localizations are listed in columns 5 and 6, and the corresponding latitude and longitude in Supergalactic coordinate systems are listed in columns 7 and 8. The conversion from celestial coordinate system to the Supergalactic coordinate system is done by using the online tools provided on a NASA website\footnote{http://lambda.gsfc.nasa.gov/toolbox/tb\_coordconv.cfm}. Column 9 lists the galaxy information within the field of the GRBs, with all the references given in column 10. All information of the galaxies are drawn from the reports on GCN Circulars and the Digital Sky Survey\footnote{http://archive.stsci.edu/cgi-bin/dss\_form}. The sky distribution of these long $\tau_{lag}$ GRBs in Supergalactic coordinates are plotted in Figure 1. At first glance, there is no tendency of concentration either towards the Supergalactic plane or towards the Virgo or Coma Cluster. More detailed analysis are presented in the next section. 

\section{Four Tests}

\subsection{Concentration Towards Supergalactic Plane}

Our Local Supercluster has a flattened distribution, with $60\%$ of its luminous galaxies in the structure which is called the plane of the Local Supercluster, and the other $40\%$ lies in five clouds off the plane, called the `halo'. If long $\tau_{lag}$ GRBs reside in galaxies in our Local Supercluster, they will show a tendency of concentration towards the Supergalactic plane. 

To quantitatively measure the tendency of the concentration, a quadruple moment of the distribution can be calculated, with $Q = <\sin^2 b - 1/3>$ and $\sigma_Q = \sqrt{4/(45N_{GRB})}$ \citep{bri96}, where \textit{b} is the latitude of GRBs in Supergalactic coordinate and $N_{GRB}$ is the number of GRBs. A significant concentration towards the plane will result in a negative Q value, while an isotropic distribution will result in a near-zero Q value. Both the quadruple moments of \citet{nor02} ($Q \sim -0.10 \pm 0.04$) and of \citet{fol08} ($Q = -0.225 \pm 0.090$) show high significance (with $|Q| \geqslant 2.5 \sigma_Q$) of a concentration towards the Supergalactic plane.

For our long $\tau_{lag}$ burst sample from \textit{Swift}, by simply counting the number of GRBs, we get only 8 of a total of 18 ($\sim44\%$) lying between -30$^{\circ}$ and 30$^{\circ}$ in Supergalactic latitude, which is in agreement with the area coverage percentage within the usual uncertainties. The calculated quadruple moment of this distribution is $Q = -0.02\pm 0.07$. It is not significantly negative. Instead, the Q value equals zero within $1-\sigma$ uncertainty and this is an indication of a homogeneous distribution. We also raised the lower limit of the `long $\tau_{lag}$' criteria to $\tau_{lag}>1.5$ s and $\tau_{lag}>2$ s, and calculated the quadruple moment for these subsamples. The results of $Q = -0.02 \pm 0.08$ for $\tau_{lag}>1.5$ s and $Q = -0.06 \pm 0.09$ for $\tau_{lag}>2$ s also show no tendency towards the Supergalactic plane. The samples and our results are shown in Table~\ref{tbl-2}. A real-time sky map of \textit{Swift} GRBs\footnote{http://grb.sonoma.edu/} shows a nearly isotropic sky distribution for \textit{Swift} bursts, and the quadruple moment for all short $\tau_{lag}$ ($\tau_{lag}<1$ s) long duration GRBs ($Q = -0.03\pm0.02$) also shows an isotropic sky distribution, with no tendency either towards or away from the Supergalactic plane. With this, we see that \textit{Swift} has a uniform sky coverage for the purpose of this paper, and so our quadruple moment of the long $\tau_{lag}$ bursts needs no correction for sky coverage. As such, we find no concentration towards the Supergalactic plane, and the Supergalactic hypothesis fails our first test.

\subsection{Concentration Towards Virgo or Coma Cluster}

The majority of the mass in our Local Supercluster is towards the Virgo Cluster and the Coma Cluster (which is in about the same direction as the Virgo Cluster, but with much larger distance from the Earth). So if these long $\tau_{lag}$ GRBs are from the Local Supercluster, there should be a tendency of concentration towards the Virgo and Coma Clusters. A dipole moment can be calculated to quantitatively measure the concentration, with $D = <\cos \theta>$ and $\sigma_D = \sqrt{1/(3N_{GRB})}$ \citep{bri96}, in which $\theta$ is the angle between the GRB and the Virgo or Coma Cluster. A concentration towards the Virgo or Coma Cluster will result in a positive dipole moment, while an isotropic distribution would result in a near-zero dipole moment. D values for a majority of long \textit{Swift} GRBs (331 bursts with $\tau_{lag}<1$ s) shows no tendency towards or away from Virgo and Coma Clusters, as shown in Table \ref{tbl-2}. The fact that the dipole for the $\tau_{lag}<1$ s bursts is closely zero tells us that the \textit{Swift} sky coverage is sufficiently uniform for the purpose of this paper and no correction to our measured D values is needed.

We calculated the dipole moment of our long $\tau_{lag}$ GRBs, towards both the Virgo and Coma Clusters. For the Virgo Cluster, the dipole moments are $-0.25 \pm 0.14$ for $\tau_{lag}>1$ s sample, $ -0.30\pm 0.15$ for the $\tau_{lag}>1.5$ s subsample, and $ -0.28 \pm 0.17$ for the $\tau_{lag}>2$ s subsample. While for Coma Cluster, the calculated dipole moments are respectively $-0.14\pm0.14$, $-0.20\pm0.15$, and $-0.18\pm0.17$ for the three cuts on $\tau_{lag}$. With the \textit{negative} dipole moments, \textit{Swift} long $\tau_{lag}$ bursts are showing a tendency \textit{away from} the Virgo and Coma clusters. Hence the hypothesis that these GRBs are from the Local Supercluster fails our second test. By checking Figure 5 in \citet{nor02} and Figure 3 in \citet{fol08}, we do not see any tendency towards the Virgo or Coma Cluster.

\subsection{Host Galaxy of these GRBs}

As long GRBs are formed by the collapsing of fast-rotating massive stars, they should be located in the star forming region of galaxies (e.g. in the spiral arms of the spiral galaxies), and these galaxies should appear within the small \textit{Swift}-XRT $90\%$ error circles. If these galaxies are members of the Local Supercluster, given the scale of the Local Supercluster ($\sim40~h^{-1}$ Mpc), they should be rather nearby, and hence relatively bright. If we adopt the R-band Schechter luminosity function with $M^\ast = -21.2$ (for a Hubble constant of 65 km s$^{-1}$ Mpc$^{-1}$; \citet{lin96}), a galaxy in our Local Supercluster with luminosity of $L^\ast/10$ will have its absolute magnitude of $M = -18.7$. This limit of $L^\ast/10$ is somewhat arbitrary, but it does include $90\%$ of the mass in a standard luminosity function. Such a galaxy on the far edge of Local Supercluster (for which we adopt a distance of $\sim56$ Mpc) will have an apparent magnitude of $m = 15.0$ or brighter. An increasing of Hubble constant from 65 km s$^{-1}$ Mpc$^{-1}$ to 72 km s$^{-1}$ Mpc$^{-1}$ will cause a slightly decreasing of the distance, and hence a brighter apparent magnitude for the threshold. As a result, if the long $\tau_{lag}$ GRBs reside in our Local Supercluster, we should be able to find their host galaxies with $m \leqslant 15$. That is, any GRB from our Local Supercluster should be immediately obvious by having its bright host galaxy in the \textit{Swift}-XRT error circle.

We checked all the GCN reports regarding to these long $\tau_{lag}$ GRBs, and all these 18 GRBs have follow up observations reported except for GRB060607B (which was too close to the Sun). Possible host galaxies are found for GRB050126 and GRB060218, with redshifts of 1.29 and 0.0331. With accurate positions reported by XRT, no galaxies are found to be within the XRT $90\%$ error circles for the remaining 16 GRBs. We also searched through the Digital Sky Survey for the fields of these GRBs, and no galaxies are found to be within the XRT error circle for all of the 18 GRBs in POSS II-F archive, the limit magnitude of which is 20.8. Hence the Supergalactic hypothesis fails this test also.

GRB060218 is a very long and smooth burst with a very long lag \citep{lia06}. An optical transient was speedily discovered with UVOT \citep{mar06} and with ROTSE \citep{qui06}. The burst position is coincident with a nearby galaxy at z = 0.0331 \citep{mir06}. Later, a supernova (SN2006aj) was found at the same position \citep{mas06, sod06}. The position is on the edge of the constellation Taurus, with $\theta=128^\circ$ to the Virgo Cluster and $\theta=125^\circ$ to the Coma Cluster. For the redshift and a Hubble constant of $H_0=72$ km s$^{-1}$ Mpc$^{-1}$, the burst is $\sim140$ Mpc distant from the Earth. This is close, but certainly outside our Local Supercluster. As such, this long $\tau_{lag}$ burst is an example of an extremely under-luminous event, but is not associated with any concentration towards the Supergalactic plane.

\subsection{Redshifts}

Given that the distance of galaxies in Local Supercluster are less than 56 Mpc from the Earth (for the Hubble constant $H_0 = 72$ km s$^{-1}$ Mpc$^{-1}$), the corresponding upper limit on the redshift is $z < 0.013$. If the long $\tau_{lag}$ GRBs are from the Local Supercluster, they should be at redshift $z<0.013$ or so. 

Of our 18 long $\tau_{lag}$ GRBs, 5 have their spectroscopic redshift reported, ranging from 1.29 to 3.08 (as listed in Table~\ref{tbl-1}). These bursts are certainly far outside the Local Supercluster. The redshift of GRB050126 is measured from the spectrum of its host galaxy, while the other 4 are all from multiple absorption lines in the optical afterglow spectra, hence these redshift values are with high confidence. The sixth GRB with a spectroscopic redshift is GRB060218, with z = 0.0331, which is also too far to be inside our Local Supercluster (see previous section). With six out of six long $\tau_{lag}$ GRBs having their spectroscopic redshift much larger than the upper limit redshift of Local Supercluster (0.013), we are very confident to make the conclusion that the Supergalactic hypothesis fails this test also. While for the remaining 12 GRBs without spectroscopic redshifts, our redshift calculated from luminosity indicators $z_{ind}$ are within the range of 0.6 to 5.0 \citep{xia08}, and the $1-\sigma$ lower limit of redshifts for all these GRBs are $z>0.5$. In summary, all these \textit{Swift} long $\tau_{lag}$ bursts are certainly outside the Local Supercluster, with 17 out of 18 at $z>0.5$. Hence the Supergalactic hypothesis fails the fourth test for all of 18 long $\tau_{lag}$ GRBs.

Moreover, if we check the whole \textit{Swift} GRB catalog (with long and short $\tau_{lag}$ values), it is easy to see that only one of all the GRBs (GRB980425) with reported spectroscopic redshift are close enough to be in our Local Supercluster. The lack of low redshift GRBs in the catalog also indicates that it is impossible to have a large fraction of long $\tau_{lag}$ of GRBs in Local Supercluster.

\section{Implications}

The Local Supercluster hypothesis strongly failed all of our four tests. Although some small fraction of long $\tau_{lag}$ GRBs can still be local (e.g. GRB980425, GRB830801), our analysis on \textit{Swift} data puts a limit of $<5\%$ on the fraction of long $\tau_{lag}$ GRBs to be in Local Supercluster.

Both the results of \citet{nor02} and \citet{fol08} show a high significance (with $|Q| \geqslant 2.5 \sigma_Q$) on the tendency of concentration towards the Supergalactic plane, which is not significantly high. Given that BATSE positions have had many selections of GRBs examined for anisotropies in many directions \citep{bri96}, with this large number of trials, we can expect that some will be significant at this level.

From our analysis, only a small fraction of long $\tau_{lag}$ GRBs (less than one out of eighteen or so) could be in the Local Supercluster. Hence, the rate of long $\tau_{lag}$ GRBs in the Local Supercluster is greatly smaller than what has been reported by \citet{nor03}, and should not be included in the calculation of LIGO's detection rate.

From Table \ref{tbl-1} we see that redshifts for these long $\tau_{lag}$ GRBs ($<z>=1.61$) are not greatly lower than for other GRBs ($<z>\sim2.3$). From the logic that long $\tau_{lag}$ corresponds with low luminosity, one might be curious as to how we can detect a $\tau_{lag}>1$ s GRB at redshift as high as $z=3$ ? GRB980425 is an example of long $\tau_{lag}$ and low redshift ($z\sim0.008$) GRB, and it is \textit{very} under-luminous (with its $\gamma$-ray peak luminosity $L = 5.5 \pm 0.7\times10^{46}~erg~s^{-1}$ according to \citet{gal98}). Of course if we put GRB980425 to the redshift of $z=3$, its luminosity distance will be increasing by a factor of $\sim730$, and it's peak flux will be decreasing by a factor of $5.3\times10^{5}$. With such a low peak flux, we will definitely not be able to detect it. However, GRB980425 is not a typical GRB. It's energy is much lower than a `normal' GRB, and it falls far below the $\tau_{lag}-L$ relation curve by a factor of several hundred. GRB980425 might represent a subclass of long GRBs with long $\tau_{lag}$, soft spectrum and low luminosity, as suggested by \citet{nor02}, but with only one example, it is unreasonable for us to take all long $\tau_{lag}$ GRBs as ultra-low luminosity bursts. 

Consider a `normal' long GRB that has $\tau_{lag}=1$ s and redshift $z=3$. Its $\tau_{lag,rest}$ in the GRB rest frame would be 0.25 s. Assuming that it fits well with the $\tau_{lag}-L$ relation from \citet{xia08},
\begin{equation}
log~L = 51.31 - 1.02*log~[\tau_{lag}(1 + z)^{-1}], 
\end{equation}
its luminosity value L would be $8.40\times10^{51}$ erg s$^{-1}$. From the concordance cosmological model, the luminosity distance $d_L$ at a given redshift is calculated by 
\begin{equation}
d_L(z) = cH_0^{-1} (1+z) \int_{0}^{z} dz' [(1+z')^3 \Omega_M + \Omega_{\Lambda}]^{-1/2}. 
\end{equation} 
with $H_0 = 72$ km s$^{-1}$ Mpc$^{-1}$, $c = 3\times10^5$ km s$^{-2}$, $\Omega_M=0.27$ and $\Omega_{\Lambda}=0.73$. The luminosity distance for $z=3$ is $d_L\sim2.5\times10^4$ Mpc. Then from the inverse square law for light, $P=L/(4\pi d_{L}^2)$, the bolometric peak flux would be $P_{bolo}=1.12\times10^{-7}$ erg cm$^{-2}$ s$^{-1}$. From the luminosity and the $E_{peak}-L$ relation \citep{xia08}
\begin{equation}
log~L = 47.73 + 1.78*log~[E_{peak}(1 + z)],
\end{equation}
a low $E_{peak}$ value $E_{peak}\sim57$ keV can be adopted. From the $E_{peak}$ and average values of the low-energy power law index $\alpha=-1.1$ and high-energy power law index $\beta=-2.2$ \citep{ban93}, peak flux value in the energy range 15 keV to 150 keV can be calculated, with the result of $P=4.93\times10^{-8}$ erg cm$^{-2}$ s$^{-1}$, which is $\sim0.54$ ph cm$^{-2}$ s$^{-1}$. It is significantly higher than the trigger threshold of \textit{Swift}. As a result, there is no doubt that we can detect long $\tau_{lag}$ GRBs at a high redshift (z=3). Indeed, the long $\tau_{lag}$ GRBs at $z>1$ are consistent with the unbroken $\tau_{lag}-L$ relation. Thus, it appears that the ultra-low luminosity `class' of bursts is quite rare (roughly fewer than one-in-nineteen), and the usual $L\propto \tau_{lag}^{-1}$ relation should be used for normal long $\tau_{lag}$ bursts with reasonable confidence.

\clearpage

\begin{figure}
\epsscale{0.80}
\plotone{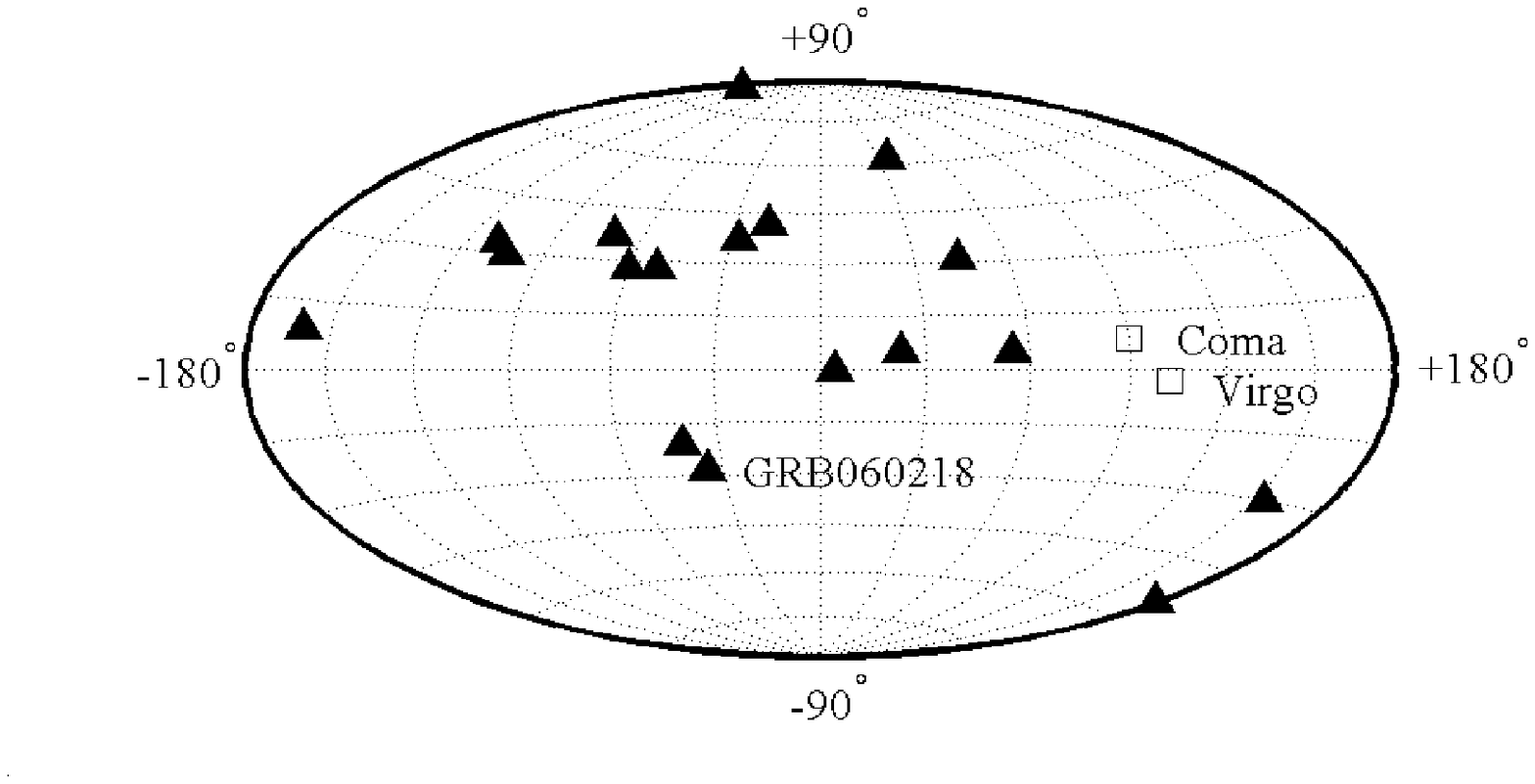}
\caption{Sky distribution of the 18 long $\tau_{lag}$ \textit{Swift} GRBs, the Virgo Cluster, and the Coma Cluster. The GRBs are marked as filled triangles, and the Virgo and Coma Clusters are marked as empty squares (upper: Coma, lower: Virgo). GRB060218, the Virgo and Coma Clusters are all marked on the right. From this figure we do not see any tendency either towards the supergalactic plane (the horizontal line running through the middle) or towards the Virgo or Coma Clusters. }
\end{figure}

\begin{deluxetable}{llllllllll}
\tabletypesize{\scriptsize}
\rotate
\tablecaption{Information of \textit{Swift} long $\tau_{lag}$ GRBs\label{tbl-1}}
\tablewidth{0pt}
\tablehead{
\colhead{GRB} & 
\colhead{$\tau_{lag}$} & 
\colhead{Redshift} & 
\colhead{Ref\tablenotemark{a}} & 
\colhead{RA\tablenotemark{b}} &
\colhead{Dec\tablenotemark{b}} &
\colhead{Longitude\tablenotemark{c}} &
\colhead{Latitude\tablenotemark{c}}		& 
\colhead{Galaxy In the Field?} & 
\colhead{Ref\tablenotemark{a}}
}
\startdata
041228	&	4.02	$\pm$	0.15	&	$2.3_{-1.2}^{+4.2}$	&	1	&	22:26:34	&	05:01:55	&	290:50:01	&	37:02:09	 &	no galaxy with $m<20.8$	&	10	\\
050126	&	2.41	$\pm$	0.08	&	1.29	&	2	&	18:32:27	&	42:23:02	&	35:37:03	&	62:53:07	 &	host z = 1.29	&	2	\\
050219A	&	2.39	$\pm$	0.17	&	$0.6_{-0.1}^{+0.2}$	&	1	&	11:05:39	&	- 40:40:51	&	154:49:21	&	-30:46:32	 &	no galaxy with $m<20.8$	&	10	\\
050410	&	3.32	$\pm$	0.30	&	$1.04_{-0.32}^{+5.15}$	&	1	&	05:59:01	&	79:36:18	&	23:56:45	&	05:36:23	 &	no galaxy with $m<20.8$	&	10	\\
050716	&	4.09	$\pm$	0.31	&	$1.4_{-0.5}^{+1.1}$	&	1	&	22:34:22	&	38:40:58	&	333:24:23	&	37:00:24	 &	no galaxy with $m<20.8$	&	10	\\
051021B	&	1.53	$\pm$	0.06	&	$2.1_{-0.7}^{+1.3}$	&	1	&	08:24:14	&	- 45:32:02	&	178:59:49	&	-54:47:23	 &	no galaxy with $m<20.8$	&	10	\\
051111	&	1.74	$\pm$	0.07	&	1.55	&	3	&	23:12:32	&	18:22:01	&	309:07:18	&	28:42:13	 &	no galaxy with $m<20.8$	&	10	\\
060218	&	177	$\pm$	16\tablenotemark{d}	&	0.03	&	4	&	03:21:31	&	16:54:36	&	325:57:54	&	-28:05:07	 &	host z = 0.0331, $M_V\sim15.8$ mag	&	4, 11	\\
060319	&	1.19	$\pm$	0.13	&	$2.0_{-0.6}^{+1.1}$	&	1	&	11:45:31	&	60:02:16	&	55:03:16	&	05:13:06	 &	no galaxy with $m<20.8$	&	10	\\
060403	&	1.15	$\pm$	0.02	&	$1.3_{-0.4}^{+0.7}$	&	1	&	18:49:21	&	08:19:37	&	195:41:24	&	82:29:38	 &	no galaxy with $m<20.8$	&	10	\\
060501	&	1.55	$\pm$	0.25	&	$1.8_{-0.5}^{+1.0}$	&	1	&	21:53:29	&	44:00:07	&	342:50:10	&	42:54:49	 &	no galaxy with $m<20.8$	&	10	\\
060502A	&	4.65	$\pm$	0.16	&	1.51	&	5	&	16:03:44	&	66:36:14	&	44:59:56	&	31:55:36	 &	no galaxy with $m<20.8$	&	10	\\
060607A	&	1.34	$\pm$	0.05	&	3.08	&	6	&	21:58:49	&	-22:29:45	&	257:13:41	&	30:59:00	 &	no galaxy with $m<20.8$	&	10	\\
060607B	&	2.98	$\pm$	0.27	&	$1.3_{-0.4}^{+0.6}$	&	1	&	02:48:10	&	14:45:18	&	319:54:06	&	-21:51:22	 &	no galaxy with $m<20.8$	&	10	\\
070330	&	2.48	$\pm$	0.07	&	$2.4_{-0.8}^{+1.5}$	&	1	&	17:58:07	&	-63:47:56	&	200:08:29	&	09:44:18	 &	no galaxy with $m<20.8$	&	10	\\
070506	&	2.28	$\pm$	0.08	&	2.31	&	7	&	23:08:48	&	10:42:39	&	300:21:59	&	28:14:04	 &	no galaxy with $m<20.8$	&	10	\\
070621	&	2.75	$\pm$	0.37	&	$1.5_{-0.4}^{+0.7}$	&	1	&	21:35:13	&	-24:48:32	&	251:04:31	&	33:43:47	 &	no galaxy with $m<21.5$	&	9	\\
071101	&	1.45	$\pm$	0.01	&	$3.7_{-1.5}^{+2.9}$	&	1	&	03:12:43	&	62:31:26	&	04:05:20	&	-00:37:40	 &	no galaxy with $m<20.8$	&	10	\\
\enddata

\tablenotetext{a}{References:-----(1)  Xiao \& Schaefer 2008;
	(2)	Berger et al. 2005;
	(3)	Hill et al. 2005;
	(4)	Mirabal et al. 2006;
	(5)	Cucchiara et al. 2006;
	(6)	Ledoux et al. 2006;
	(7)	Thoene et al. 2007;
	(8)	Jakobsson et al. 2008;
	(9)	Bloom et al. 2007;
	(10)  Digital Sky Survey: http://archive.stsci.edu/cgi-bin/dss\_form
	(11)	Modjaz et al. 2006; }

\tablenotetext{b}{The Right scension and declination values are in the celestial coordinate system.}
\tablenotetext{c}{The longitude and latitude values are in the supergalactic coordinate system.}
\tablenotetext{d}{Value obtained from \citet{lia06}.}
\end{deluxetable}

\begin{deluxetable}{lllll}
\tabletypesize{\scriptsize}
\center
\tablecaption{Dipole and Quadruple Statistics\label{tbl-2}}
\tablewidth{0pt}
\tablehead{
\colhead{Sample} & 
\colhead{$N_{GRB}$} & 
\colhead{Q\tablenotemark{a}}  &
\colhead{D for Virgo\tablenotemark{b}} & 
\colhead{D for Coma\tablenotemark{b}}
}
\startdata
$\tau_{lag}<1$ s	&	331	&	-0.03		$\pm$	0.02	&	-0.03		$\pm$	0.03	&	-0.01		$\pm$	0.03		\\
$\tau_{lag}>1$ s	&	18	&	-0.02		$\pm$	0.07	&	-0.25 	$\pm$ 	0.14	&	-0.14		$\pm$	0.14		\\
$\tau_{lag}>1.5$ s	&	14	&	-0.02 	$\pm$	0.08	&	-0.30 	$\pm$	0.15	&	-0.20		$\pm$	0.15		\\
$\tau_{lag}>2$ s	&	11	&	-0.06		$\pm$	0.09	&	-0.28 	$\pm$	0.17	&	-0.18		$\pm$	0.17		\\
\enddata

\tablenotetext{a}{The quadruple moment is sensitive to measuring a concentration towards the Supergalactic plane ($Q\ll0$), while an isotropic distribution yields $Q\simeq0$.}
\tablenotetext{b}{The dipole moment is sensitive to measuring a concentration towards either the Virgo Cluster or the Coma Cluster ($D\gg0$), while an isotropic distribution yields $D\simeq0$.}

\end{deluxetable}


\begin{thebibliography}{}
\bibitem[Band et al.(1993)]{ban93} Band, D. et al. 1993, ApJ, 413, 281.
\bibitem[Band (1997)]{ban97} Band, D. L. 1997, ApJ, 486, 928.
\bibitem[Berger et al.(2005)]{ber05}	Berger, E. et al. 2005, GCN 3088.
\bibitem[Bloom et al.(2007)]{blo07} Bloom, J. S., Chen, H.-W., Perley, D. A. \& Pollack, L. 2007, GCN 6568. 
\bibitem[Briggs et al.(1996)]{bri96} Briggs, M. S. et al. 1996, \apj, 459, 40.
\bibitem[Cucchiara et al.(2006)]{cuc06}	Cucchiara, A. et al. 2006, GCN 5052.
\bibitem[de Vaucouleurs (1953)]{dev53} de Vaucouleurs, G. 1953, \aj, 58, 30.
\bibitem[de Vaucouleurs (1958)]{dev58} de Vaucouleurs, G. 1958, \aj, 63, 253.
\bibitem[Foley et al.(2008)]{fol08} Foley, S., McGlynn, S., Hanlon, L., McBreen, S. \& McBreen, B. 2008, A \& A, 484, 143.
\bibitem[Galama et al.(1998)]{gal98} Galama, T. J., et al. 1998, Nature, 395, 670.
\bibitem[Gehrels et al.(2004)]{geh04} Gehrels, N. et al. 2004, \apj, 611, 1005.
\bibitem[Hill et al.(2005)]{hil05}	Hill, G., Prochaska, J. X., Fox, D., Schaefer, B. \& Reed, M.,  2005, GCN 4255.
\bibitem[Karachentsev \& Makarov (1996)]{kar96} Karachentsev, I. D. \& Makarov, D. A. 1996, \aj, 111, 794.
\bibitem[Lahav et al.(2000)]{lah00} Lahav, O., Santiago, B. X., Webster, A. M., Strauss, M. A., Davis, M., Dressler, A., \& Huchra, J. P. 2000, MNRAS, 312, 166.
\bibitem[Ledoux et al.(2005)]{led05} Ledoux, C. et al. 2006. GCN 5247.
\bibitem[Liang et al.(2006)]{lia06} Liang, E.-W., Zhang, B.-B., Stamatikos, M., Zhang, B., Norris, J., Gehrels, N., Zhang, J. \& Dai, Z.-G., 2006, \apj, 653, L81.
\bibitem[Lin et al.(1996)]{lin96} Lin, H. et al. 1996, \apj, 464, 60.
\bibitem[Marshall et al.(2006)]{mar06} Marshall, F. et al. 2006, GCN 4779.
\bibitem[Masetti et al.(2006)]{mas06} Masetti, N. et al. 2008, GCN 4803.
\bibitem[Mirabal et al.(2006)]{mir06} Mirabal, N. et al. 2006, GCN 4792.
\bibitem[Modjaz et al.(2006)]{mod06} Modjaz, M. et al. 2006, \apj, 645, L21.
\bibitem[Norris et al.(1986)]{nor86} Norris, J. P., Share, G. H., Messina, D. C., Matz, M., Kouveliotou, C., Dennis, B. R., Desai, U. D. \& Cline, T. L. 1986, Adv. Space Rev., 6, 19.
\bibitem[Norris et al.(2000)]{nor00} Norris, J. P., Marani, G. F., \& Bonnell, J. T. 2000, \apj, 534, 248.
\bibitem[Norris (2002)]{nor02} Norris, J. P. 2002, \apj, 579, 386.
\bibitem[Norris (2003)]{nor03} Norris, J. P. 2003, in AIP Conf. Ser. 686, The Astrophysics of Gravitational Wave Sources, ed. J. M. Centrella (College Park: AIP), 74.
\bibitem[Quimby et al.(2006)]{qui06} Quimby, R. et al. 2006, GCN 4782.
\bibitem[Schaefer et al.(2001)]{sch01} Schaefer, B. E., Deng, M. \& Band, D. L. 2001, \apj, 563, L123.
\bibitem[Schaefer(2003)]{sch03} Schaefer, B. E. 2003, \apj, 583, L67.
\bibitem[Schaefer(2004)]{sch04} Schaefer, B. E. 2004, \apj, 602, 306.
\bibitem[Soderberg et al.(2006)]{sod06} Soderberg, A., Berger, E. \& Schmidt, B, 2006, IAUC, 6784.
\bibitem[Thoene et al.(2007)]{tho07} Thoene, C. C., Jaunsen, A. O., Fynbo, J. P. U., Jakobsson, P. \& Vreeswijk, P. M. 2007, GCN 6379. 
\bibitem[Tully (1982)]{tul82} Tully R. B. 1982, \apj, 257, 389.
\bibitem[Tully \& Fisher (1987)]{tul87} Tully, R. B. \& Fisher, J. R. 1987, Atlas of Nearby Galaxies (Annales de Geophysique).
\bibitem[Uemura et al.(2007)]{uem07} Uemura, M. et al. 2007, GCN 7037.
\bibitem[Vianello et al.(2008)]{via08} Vianello, G., G\"otz, D. \& Mereghetti, S. 2008, astro-ph/0812.3349.
\bibitem[Xiao \& Schaefer(2008)]{xia08} Xiao, L. \& Schaefer, B. E., 2008, \apj, submitted.

\end{thebibliography}
\end{document}